\documentclass{article} 
\usepackage[english]{babel}
\usepackage{lmodern}
\usepackage[latin9]{inputenc}
\usepackage{endnotes}
\usepackage{amssymb,amsmath,amsfonts}
\usepackage{textcomp}
\usepackage{authblk}

\usepackage{graphicx}
\usepackage{dcolumn}
\usepackage{bm}
\usepackage{xcolor}

\begin{document}

\title{Micro-vorticity fluctuations affect the structure of \textcolor{black}{thin fluid films}}

\author{A. Tiribocchi}
\affil{Center for Life Nano Science@La Sapienza, Istituto Italiano di Tecnologia, 00161 Roma, Italy}
\affil{Istituto per le Applicazioni del Calcolo CNR, via dei Taurini 19, Rome, Italy}
\author{A. Montessori}
\affil{Istituto per le Applicazioni del Calcolo CNR, via dei Taurini 19, Rome, Italy}
\author{S. Miliani}
\affil{Department of Engineering, Roma Tre University, Via Vito Volterra 62, Rome 00146, Italy}
\author{M. Lauricella}
\affil{Istituto per le Applicazioni del Calcolo CNR, via dei Taurini 19, Rome, Italy}
\author{M. La Rocca}
\affil{Department of Engineering, Roma Tre University, Via Vito Volterra 62, Rome 00146, Italy}
\author{S. Succi}
\affil{Center for Life Nano Science@La Sapienza, Istituto Italiano di Tecnologia, 00161 Roma, Italy}
\affil{Istituto per le Applicazioni del Calcolo CNR, via dei Taurini 19, Rome, Italy}
\affil{Institute for Applied Computational Science, John A. Paulson School of Engineering and Applied Sciences, Harvard University, Cambridge, USA}

\date{\today}

\maketitle

\begin{abstract}

The dynamic interaction of complex fluid interfaces is highly sensitive to 
near-contact interactions occurring at the scale of ten of nanometers. 
Such interactions are difficult to analyse because they couple self-consistently to the 
dynamic morphology of the evolving interface, as well as to the hydrodynamics of the interstitial fluid film. 
In this work, we show that, above a given \textcolor{black}{magnitude} threshold, near-contact interactions trigger 
non-trivial micro-vorticity patterns, which in turn affect the effective near-contact 
interactions, giving rise to persistent \textcolor{black}{fluctuating} ripples at the fluid interface. 
In such regime, near-contact interactions may significantly affect the macroscopic arrangement of \textcolor{black}{emulsion} configurations,
such as those arising in soft-flowing \textcolor{black}{microfluidic} crystals.
\end{abstract}

\textcolor{black}{\section{Introduction}}

The ability to manufacture, control and manipulate soft matter systems, such as foams and emulsions, depends 
crucially on a thorough understanding of the complex dynamic interactions between fluid interfaces \cite{fernandez2016,piazza2012}. 

Such intricate dynamics is controlled by the competition of multiple concurrent interactions, including viscous 
dissipation, long-range hydrodynamics and near-contact forces \cite{oettinger1995,bergeron1999}. 
The latter, in particular, emerge slightly above the nanometer scale, \textcolor{black}{which is the typical
thickness of the fluid film separating fluid droplets,} and dominate the dynamics \cite{prosperetti2007,kruger2013}.

A remarkable example in point is offered by  {\it soft flowing crystals} \cite{garstecki2006flowing,montessori2019mesoscale},
soft porous materials manufactured via microfluidic devices (such as flow focusers) and of great potential interest
in applications like catalyst support \cite{kimmins2010,costantini2014} or scaffolds in tissue engineering \cite{li2018microfluidic,hamley2009ordering,costantini2015micro,marmottant2009,marmottantprl}. 

In these systems, near-contact interactions (NCIs), such as disjoining pressure and dispersion forces 
\cite{stubenrauch2003disjoining,stone1996}, are triggered in the intervening thin film of fluid separating the droplets, due to 
the close contact of different portions of fluid interfaces, and typically prevent droplet coalescence \cite{chan2011film,dagastine2006}. 

Despite their short-range nature, the effect of these forces may span across several length scales. 
For instance, the interaction between sub-millimetre size oil drops across thin films of a continuous water phase 
may produce interface deformations, extending from tens to hundred nanometres \cite{chan2011film}. 
This may, in turn, lead to long-range rearrangements of the structure of the emulsion 
(i.e. plastic events \cite{cantat2013,cheddadi2013}), often propagating up to the millimeter lengthscale \cite{sanfeld2008}, 
thereby affecting the stability of the material. 

Therefore, understanding the complex nature of these near-contact forces is a fundamental step in the direction of 
an accurate design of soft porous materials with enhanced performances. 

In spite of the enormous progress in the understanding of the role played by such forces \cite{derjaguin1940,derjaguin1941,verwey1948,webber2008,dagastine2006,chan2011film,wang2015,montessori2019mesoscale}, a firm picture of the 
way that the microscopic details of NCI's affect the large-scale dynamics of a soft porous material, it is still missing.
For instance, how do thin film deformations (i.e. ripples) propagate along the fluid interface? 
And, what is the interplay between such deformations and the fluid flow within the film?

From a theoretical standpoint, such questions can be tackled almost exclusively by numerical simulations, due 
to the complex structure of the physical equations, which rules out analytical solutions for all but a few
precious exceptions \cite{degennes2004}. 

On the other hand, a full-scale numerical investigation of the physics of soft flowing systems, is still a major 
challenge, since including near-contact forces at the interface level requires the concurrent handling
of five-six spatial decades, from millimetres (the typical size of a microfluidic device) all the way 
down to nanometers, the relevant scale of contact forces. 

This issue naturally calls into play the concept of universality, namely the 
extent to which suitable dimensionless parameters measuring the relative strength of NCIs versus, say, capillary forces, 
prove capable of capturing the relevant features of the underlying physics. 

If such universality holds, one can resort to a less computationally demanding mesoscale approach 
to describe the  effects of these forces on the dynamics of complex flowing systems, with no need 
of explicitly including  the details of the molecular interactions. 

This is precisely the idea behind the present paper.

Here we numerically investigate the near-contact physics observed within a thin film formed during 
the impact of two micron-sized fluid droplets, using a recent extension of the mesoscale lattice Boltzmann method \cite{succi2018lattice,kruger2017lattice,montessori2014regularized,montessori2015lattice}. 

The effects of near-contact forces are modelled via a mesoscopic repulsive force,
which competes with capillary forces (surface tension) to prevent 
droplet coalescence and ensuing coarsening of the material \cite{montessori2019jfm}. 

We find that above a given \textcolor{black}{magnitude} threshold, the near-contact forces trigger \textcolor{black}{local} micro-vorticity
recirculation patterns which, in turn, excite interface ripples exhibiting persistent \textcolor{black}{fluctuations}. 
Such strongly non-equilibrium phenomena emerge whenever the repulsive forces become comparable
or exceed the effects of surface tension, a competition which is measured by the
dimensionless number ${\cal N}_c=A\Delta x/\sigma\geq 1$, where $A$ gauges the strength of 
the repulsive force within the lattice spacing $\Delta x$ (characteristic lengthscale of NCIs) and $\sigma$ is the surface tension. 

The typical height \textcolor{black}{$h$} of the liquid film at which these events occur matches to a \textcolor{black}{reasonable} accuracy  
the experimental values  reported in Ref. \cite{klaseboer2000}.
Thanks to its ability to fully resolve the hydrodynamic structure of the flow within the film, our numerical model reveals 
the presence of  micro-vortices and unidirectional fluxes, which develop laterally and in its middle, 
thereby promoting fluctuating interface undulations.

\textcolor{black}{The paper is organized as follows. In Section II we illustrate the details of the simulation approach while
Section III is dedicated to describe the numerical results. In particular Section IIIA elucidates the thin film dynamics and the interface fluctuations, and Section IIIB
focuses on the hydrodynamic effects and on the role of the vorticity in the system. Finally, Section IV concludes the manuscript.}

\textcolor{black}{\section{Method}}

\textcolor{black}{The computational model used in this work  is based on a recent development of a Lattice Boltzmann (LB) approach for multicomponent flows \cite{montessori2019jfm,leclaire2012,succi2018lattice}, aimed at capturing the effects of near-contact interactions operating at the fluid interface level. Such method has been found to correctly reproduce the collision between bouncing droplets \cite{montessori2019jfm} and to simulate the dynamics of soft flowing crystals in a flow focuser \cite{montessori2019mesoscale}.}

\textcolor{black}{\subsection{Numerical model and equations of motion}}

\textcolor{black}{The method is built starting from two sets of distribution functions tracking the evolution of the two fluid components, i.e. the droplet phase and the outer surrounding fluid. Their dynamics is governed by a discrete Boltzmann equation of the form:}

\textcolor{black}{\begin{equation} \label{CGLBE}
f_{i}^{k} \left({\bf x}+{\bf c}_{i}\Delta t,\,t+\Delta t\right) =f_{i}^{k}\left({\bf x},\,t\right)+\Omega_{i}^{k}( f_{i}^{k}\left({\bf x},\,t\right)),
\end{equation}
where $f_{i}^{k}$ represents the probability of the $k^{th}$ component of finding a fluid particle at time $t$ at position ${\bf x}$ moving with a discrete velocity ${\bf c}_{i}$. The index $i$ runs over the lattice discrete directions $i = 0,...,N$, where $N=26$, i.e. we are considering a three dimensional lattice with $27$ lattice velocities. As common in LB simulations \cite{succi2018lattice}, we take the time step $\Delta t=1$.}

\textcolor{black}{The total fluid density and the total momentum of the mixture are given by
\begin{equation}
\rho=\sum_k\rho^k, \hspace{1cm} \rho {\bf v} = \sum_k  \sum_i f_{i}^{k}\left({\bf x},\,t\right){\bf c}_{i},
\end{equation}
where $\rho^{k}({\bf x},t) = \sum_i f_{i}^{k}({\bf x},t)$.}

\textcolor{black}{The collision operator $\Omega_{i}^{k}$ consists of three terms \cite{gunstensen1991,leclaire2012,leclaire2017}, and is given by
\begin{equation}
\Omega_{i}^{k} = \left(\Omega_{i}^{k}\right)^{(3)}\left[\left(\Omega_{i}^{k}\right)^{(1)}+\left(\Omega_{i}^{k}\right)^{(2)}\right].
\end{equation}}

\textcolor{black}{Here $\left(\Omega_{i}^{k}\right)^{(1)}$ is the relaxation step (\cite{succi2018lattice})
and is related to the kinematic viscosity $\nu$ of the mixture.
The second term $\left(\Omega_{i}^{k}\right)^{(2)}$ is the perturbation step (\cite{gunstensen1991}) and models the interface tension between the two components, while the third term $\left(\Omega_{i}^{k}\right)^{(3)}$ is the recoloring step (\cite{gunstensen1991,latva2005}) and enforces the separation between the two fluid.}

\textcolor{black}{It can be shown that, by performing a Chapman-Enskog expansion of the distribution functions and of their derivatives \cite{chen1998,benzi1992}, continuity and Navier-Stokes equations can be recovered in the continuum limit
\begin{equation}
\frac{\partial \rho}{\partial t} + \nabla \cdot {\rho {\bf v}}=0 
\end{equation}
\begin{equation}\label{NSE}
\frac{\partial \rho {\bf v}}{\partial t} + \nabla \cdot {\rho {\bf v}{\bf v}}=-\nabla p + \nabla \cdot [\rho \nu (\nabla {\bf v} + \nabla {\bf v}^T)] + \nabla \cdot \boldsymbol{\Sigma}.
\end{equation}
In Eq.\ref{NSE}, $p$ is the ideal gas pressure and $\boldsymbol{\Sigma}$ is the stress tensor, a term that captures the effect of the interface tension.}

\textcolor{black}{Finally one needs to specify the boundary conditions of the velocity field at the wall, where one of the two droplets is attached. We have set no-slip conditions, meaning that ${\bf v}_w=0$.}

\begin{figure}
\includegraphics[width=0.7\linewidth]{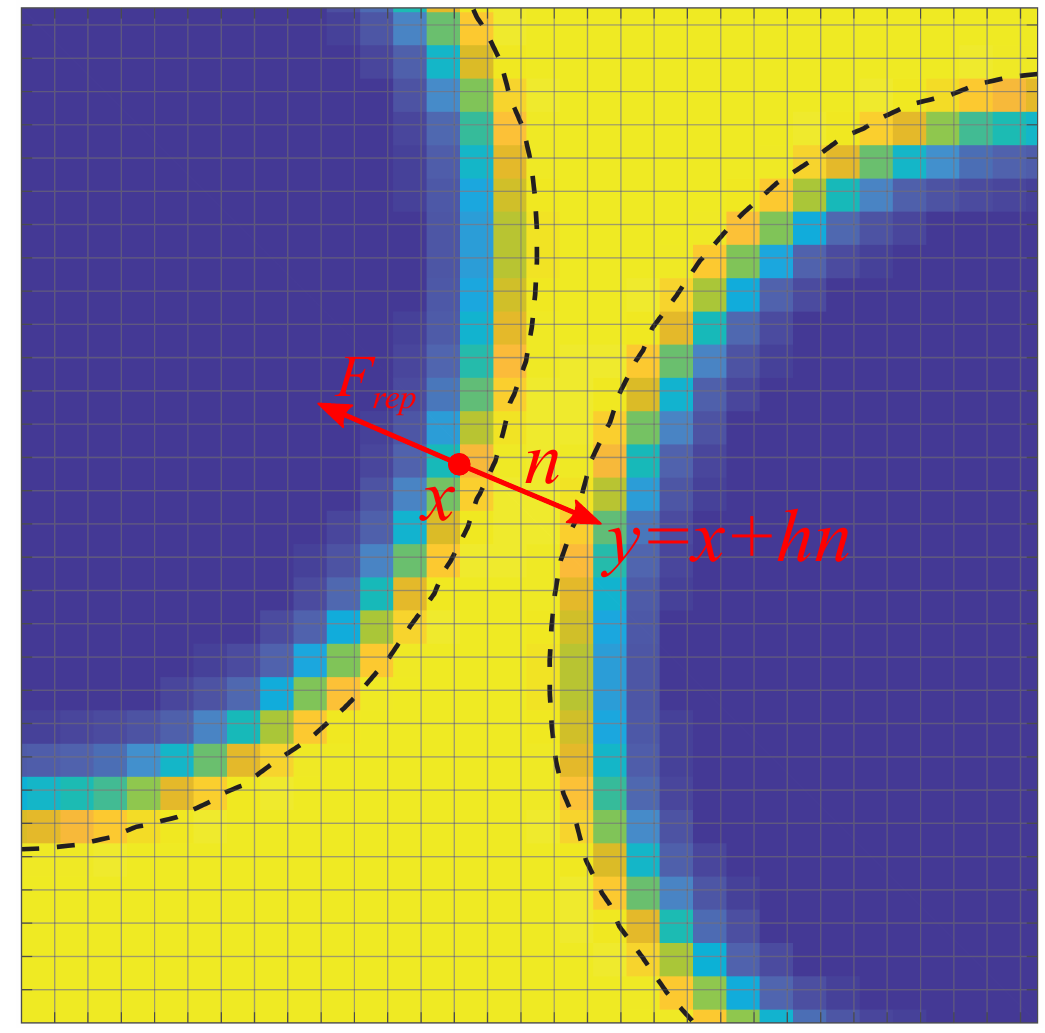}
\caption{\textcolor{black}{Mesoscale modelling of near-contact interactions between two immiscible fluid droplets. ${\bf F}_{rep}$ represents the repulsive force and ${\bf n}$ is the unit vector perpendicular to the fluid interface. ${\bf x}$ and ${\bf y}$ indicate the positions, placed at distance $h$, taken within the fluid interface, and the dotted line tracks its outermost frontier.}}
\label{figS1}
\end{figure}

\textcolor{black}{A mesoscale representation of the near-contact forces (including Van del Waals, electrostatic and steric forces) can be achieved by adding a further term to the stress tensor $\boldsymbol{\Sigma}$ accounting for their repulsive effect. Such repulsive force can be written as
\begin{equation}\label{Frep}
{\bf F}_{rep}\equiv\nabla\pi=-A_h[h({\bf x})]{\bf n}\delta_I,
\end{equation}
where $h({\bf x})$ is the distance between positions ${\bf x}$ and ${\bf y}={\bf x}+h{\bf n}$ located at the two interfaces, and ${\bf n}$ is the normal direction, perpendicular to the interface and pointing outwards the droplet phase (see Fig.\ref{figS1}). The term $A_h[h({\bf x})]$ sets the strength of the near-contact forces and is a ``coarse-grained'' analogous of the Hamaker constant from the DLVO theory \cite{derjaguin1940,derjaguin1941,verwey1948}. It is constant and equal to $A$ if $h<h_{min}$, while it decays as $h^{-3}$ if $h>h_{min}$. 
Finally $\delta_I=\frac{1}{2}\nabla\phi$ is a function confining the near-contact interactions at the fluid interface.}

\textcolor{black}{This additional contribution (localized at the interaface) modifies the total stress of the Navier-Stokes equation, which can be recovered with the formal substitution $\nabla\cdot \boldsymbol{\Sigma}\rightarrow \nabla\cdot(\boldsymbol{\Sigma}+\pi\boldsymbol{I})$, being $\boldsymbol{I}$ the unit matrix.}

\textcolor{black}{\subsection{Simulation details and numerical mapping}}

\textcolor{black}{Shape deformations of the droplet can be described by the Capillary number $Ca=\rho\nu V/\sigma$ and by the Bond number $Bo=\Delta\rho g (2R)^2/\sigma$. Here $V$ is a characteristic velocity of the droplet, $\Delta\rho$ is the density difference between the two phases and $g$ is the gravitational acceleration. The Capillary number gauges the  strength of the viscous forces relative to surface tension at the droplet interface, whereas the Bond number measures the effect of gravitational forces with respect to $\sigma$. In our simulations we have $Ca\sim 10^{-2}$ and $Bo\sim 10^{-1}$, i.e. both viscous and gravitational forces are much smaller than the surface tension.}

\textcolor{black}{Simulations are performed on a quasi-2d lattice of size $L_x=600$, $L_y=400$, $L_z=3$ in which two equally sized micro-droplets of fluid are immersed in the bulk of a second immiscible fluid. One droplet  is attached at the bottom wall, while the other is left free to fall under the effect of a body force ${\bf G}$ applied along the $y$-direction (perpendicular to the wall), which mimics the gravitational force.}

\textcolor{black}{Unless otherwise stated, parameter values are: $\Delta t=1$, $\Delta x=1$, $A\sim 0.05 - 0.5$, $h_{min}=4\Delta x$, $R\sim 60$, $Ca\sim 10^{-2}$, $\mu=0.167$, $\rho=1$, $g=10^{-5}$. An approximate mapping between simulation units and physical ones can be obtained by comparing our results with a typical collision experiment of near millimetric droplets of immiscible fluids \cite{chen2006}. By taking the lattice spacing $\Delta x$ corresponding to approximately $1\mu$m, one obtains a droplet of diameter $\sim 100\mu$m, and an impact velocity within $1-3m/s$. Within such scales, the thickness $h$ of the film of fluid ranges between $3-4\mu$m, approximately three orders of magnitude larger than the characteristic experimental values, of the order of nanometers and unaccessible to our simulations. Nonetheless, this result, once more, brings into play the concept of universality of a physical process, which states that the interaction physics, at the spatial scale considered, can be fully captured by dimensionless numbers, such as $Ca$ and ${\cal N}_{c}$, rather than by the absolute strength of the interactions.}

\textcolor{black}{\section{Results}}

\begin{center}
\begin{figure*}
\includegraphics[width=0.8\linewidth]{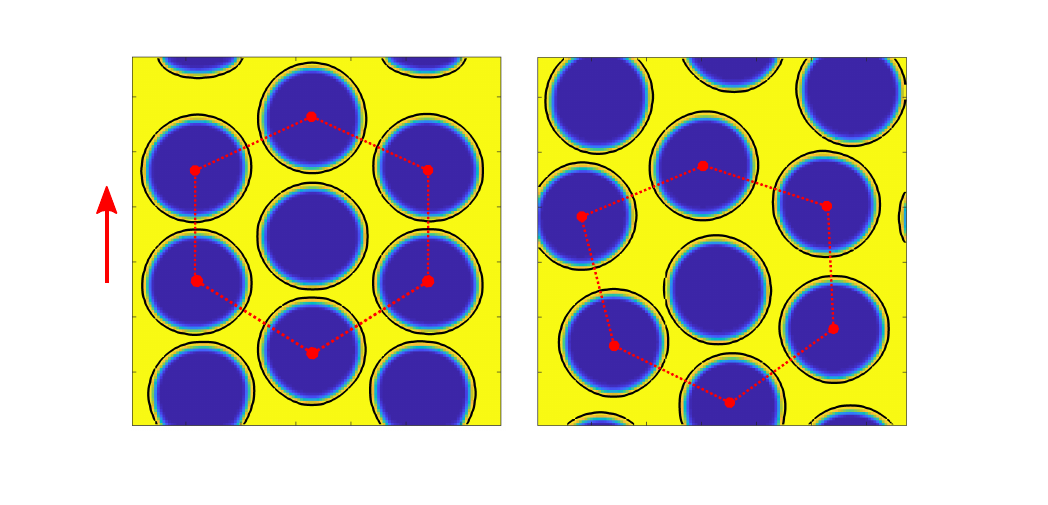}
\caption{Mono-disperse dense emulsion in a microfluidic channel under an external constant body force.
\textcolor{black}{The red arrow indicates its direction}. (Left) 
\textcolor{black}{If near-contact forces are weak (${\cal N}_c\sim 0.1$), an ordered pattern of droplets exhibiting a hexagonal
crystal-like order is produced. The soft crystal flows without significant structural deformations.}
(Right) \textcolor{black}{If near-contact forces are high enough (${\cal N}_c\sim 4$), the regular crystalline texture is deformed.}
Dotted lines connect the droplet centers of mass, thus   providing a visual picture at the dynamic connectivity of the
macroscopic pattern. \textcolor{black}{Only a portion of the micro-channel is shown.}}
\label{fig1} 
\end{figure*}
\end{center}

The present study is meant to inspect the fundamentals of the near-contact two-body (droplet-droplet) physics 
in order to set the stage for more complex applications, involving larger collections of droplets, the
main question being whether and to what extent do the interface undulations affect the macroscopic pattern 
of \textcolor{black}{ordered emulsion} configurations, such as that shown in Fig.\ref{fig1}. Here, a crystal-like structure is produced
in a microfluidic channel by a highly ordered mono-disperse emulsion \textcolor{black}{made of an immiscible binary fluid mixture (left).
Similar structures have been experimentally realized, for instance, in Ref.\cite{costantini2014,marmottant2009,marmottantprl}.
If near-contact interactions are lower than the surface tension (${\cal N}_c<1$), the hexagonal crystalline texture remains unaltered. If, on the contrary,
NCIs are comparable of higher than $\sigma$ (${\cal N}_{c}>1$), the soft crystal is deformed (right).
Hence, a careful balance of the effect produced by the NCIs is essential for controlling shape and morfphology of a soft flowing crystal.}

\begin{center}
\begin{figure*}
\includegraphics[width=0.7\linewidth]{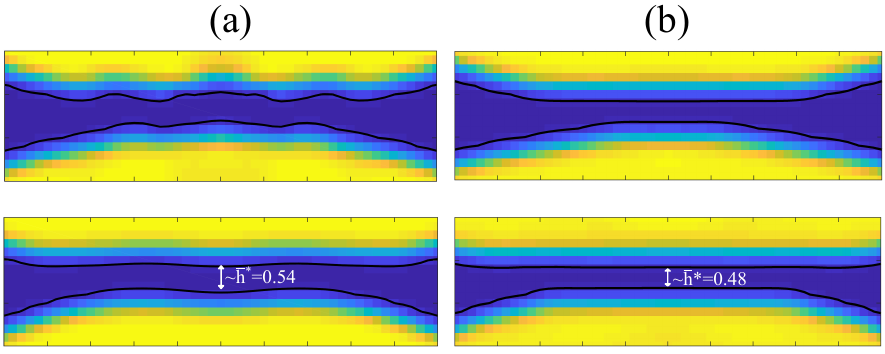}
\caption{(a)-(b) Time evolution of the dynamics of thin film (in blue) formed between two fluid droplets (in yellow). 
The black line tracks the position of the interface. 
(a) Top: If ${\cal N}_c>1$, interface ripples form and propagate through the film of fluid. 
Bottom: Interfaces return approximately smooth until a new ripple form. 
The dimensionless interface distance $h^{*}$ fluctuates in time. 
(b) If ${\cal N}_c<1$ no ripples occur, although near-contact forces are sufficiently intense to prevent droplet coalescence. 
Fluid interfaces look roughly smooth and $h^{*}$ remains constant with time. 
}\label{densityprofcomp}
\end{figure*}
\end{center}
\begin{center}
\begin{figure}
\includegraphics[width=0.8\linewidth]{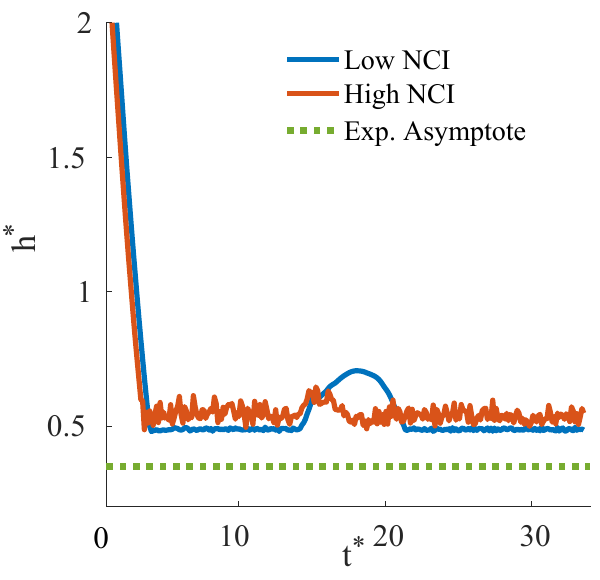}
\caption{Time evolution of the film thickness $h^*$ for a high value (orange line) and a low value (blue line) of NCI.
The local maximum around $t^*\sim 18$ (low NCI) denotes a temporary rebound of the two droplets. \textcolor{black}{Persistent fluctuations emerge at late times for high NCIs.}
Dotted line is the experimental asymptotic value taken from Ref.\cite{klaseboer2000}. }
\label{hfluctuations} 
\end{figure}
\end{center}

\textcolor{black}{\subsection{Thin film dynamics and interface fluctuations}}

In Fig.\ref{densityprofcomp}, we show a time sequence of the thin film dynamics between the droplets. 
The droplet on the top falls under the effect of the body force ${\bf G}$, moving towards the one anchored to the bottom wall. 
As long as droplets are far apart, their interfaces remain approximately circular, a picture 
which changes drastically once they come in near contact (Fig.\ref{densityprofcomp}a). 
Interestingly, whenever ${\cal N}_c>1$, the thin film between the interfaces starts to deform, and ripples spontaneously emerge. 
Such interface deformations occur because, during the thinning process, the near-contact forces withstand 
the combined effect of gravity and surface tension, which would favour coalescence and decrease of interfacial area. 

\begin{center}
\begin{figure*}
\includegraphics[width=0.8\linewidth]{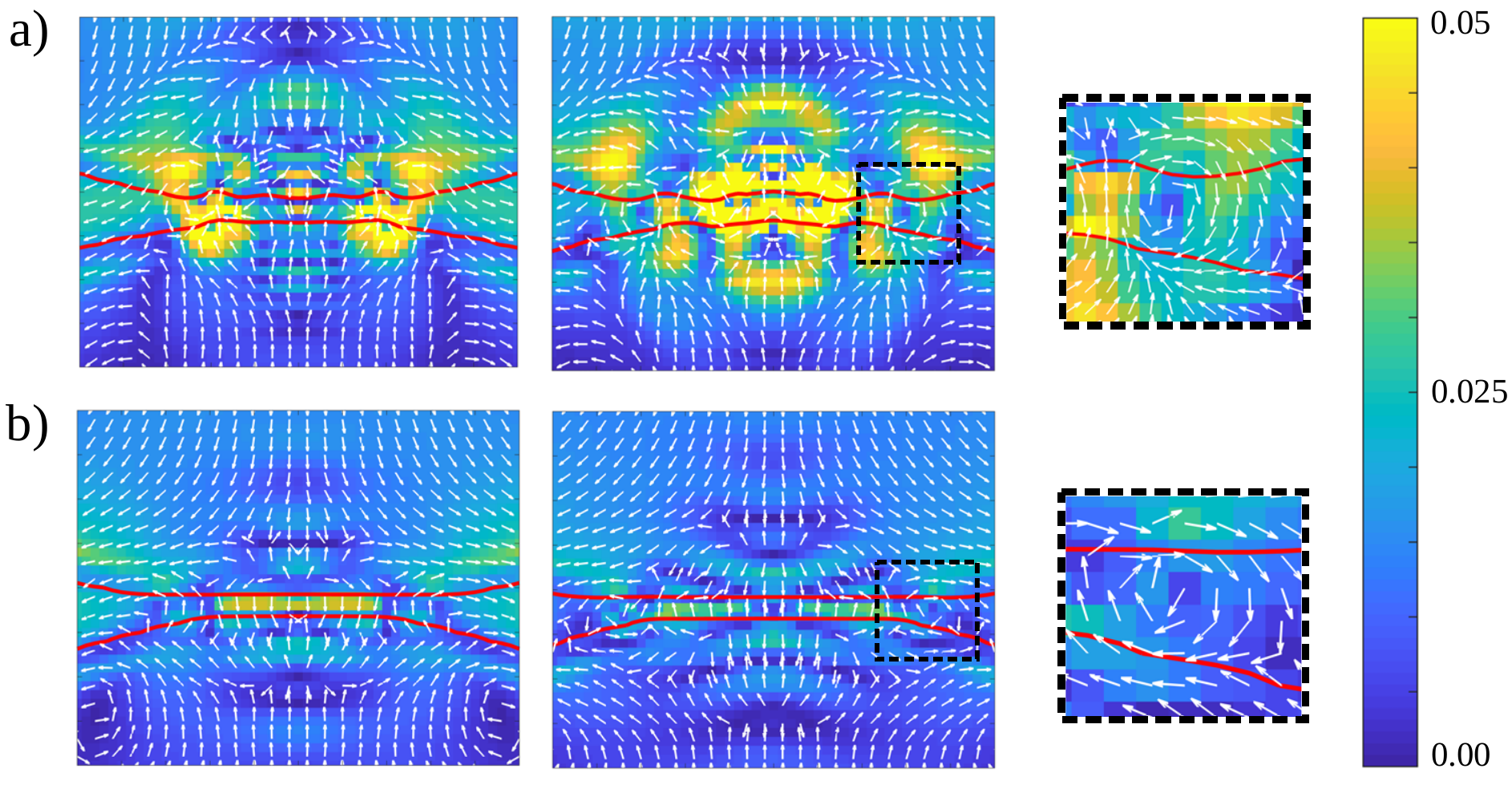}
\caption{Hydrodynamic flow field within the film. Fluid flow within the film and in its surroundings for (a) ${\cal N}_{c}>1$
 and (b) ${\cal N}_{c}<1$ \textcolor{black}{taken at $t^*\simeq 6.6$ (left column) and $t^*\simeq 8$ (right column)}. 
White vectors indicate the local fluid direction, while the color map sets its magnitude (in simulation units). 
(a) Large flows emerge within the intervening film, where two dynamic counter-rotating vortices located at 
its extremities, push the fluid outwards, surrounding a chaotic-like fluid flow in the middle. 
Ripples are observed mainly in correspondence with the most intense flow events. 
(b) A more regular fluid flow pattern is found. 
Weaker lateral vortices encircle a uniform fluid flow in the middle of the film. 
The zooms show a close view of the fine-scale spatial structure of the vortices.}
\label{flowfield}
\end{figure*}
\end{center}

\textcolor{black}{When this occurs,
micro-recirculating patterns emerge and trigger the formation of ripples propagating along the interface from the  centre of the film to its periphery,
simultaneously exhibiting persistent fluctuations. Micro-recirculations typically result from the structure of the fluid flow, entering from the top and
recirculating away from the bottom to sustain the film in time (see Fig.\ref{flowfield}).}

Subsequently, the interface partially recovers its smooth profile, before a new ripple forms again, thus initiating a new cycle. 

This fascinating phenomenon has been experimentally observed in the drainage of thin films in Scheludko cells, in which 
a dimple spontaneously forms, grows around its the centre and then leaves a parallel film behind it \cite{scheludko1967,exerowa1997}. 
Such process occurs cyclically and can persist for hours, sustained by the redistribution of the surfactant at the 
interface between the phases \cite{velev1993}. 

A quantitative signature of this complex motion is captured by inspecting the time evolution of the 
non-dimensional height $h^{*}$ of the liquid film. 

In Fig.\ref{hfluctuations}, we plot $h^{*}=h/h_r$, computed on a vertical line passing through 
the centres of the droplets, versus the non-dimensional time $t^{*}=t/t_r$,  for both ${\cal N}_c>1$ 
(high NCIs) and ${\cal N}_c<1$ (low NCIs). 
Here $h$ and $t$ are the values of interface separation and time of simulations, 
while $h_r=Ca^{1/2}R$ \textcolor{black}{is the typical thickness of the film estimated via dimensional analysis of the thin film equations \cite{klaseboer2000}}, and 
$t_r=2 \pi R/ V_c$ \textcolor{black}{is the typical time a capillary wave, moving at speed $V_c$, takes to cover the perimeter of a droplet of radius $R$. Finally,} 
 $V_c=\sigma/\mu$ is the capillar velocity and $\mu$ is the dynamic viscosity. 
For the current simulation, $R \sim 60$, $t_r \sim 400$, both in lattice units and $Ca \sim 10^{-2}$.

Persistent \textcolor{black}{fluctuations} of $h^*$ are triggered whenever ${\bf F}_{rep}$ is sufficiently intense, and are 
not observed when ${\bf F}_{rep}$ is lowered by a factor $\sim 10$ (see Fig.\ref{densityprofcomp}b, where the fluid film 
looks nearly flat), even though droplet coalescence is still inhibited. 
The late-time average value of $h^*$  at low values of ${\bf F}_{rep}$ remains approximately constant 
at $0.48$ throughout the simulation, while the one calculated for larger values of ${\bf F}_{rep}$ fluctuates around $0.54$, not far 
from the experimental findings \cite{klaseboer2000,karakashev2010}, which report ${h}^*\sim 0.35$. 
Given the massive coarse-graining inherent to our model, basically two-three spatial 
decades, this can be regarded as a very encouraging result. 
A possible source of discrepancy may be attributed to a partial lack of universality, or simply 
to a non-optimal calibration of our mesoscopic model.

\textcolor{black}{These results support the view that}
the transition to the unsteady fluctuating behaviour is controlled by ${\cal N}_{c}$. \textcolor{black}{Indeed, three different dynamic regimes can be observed:} 
as long as the surface tension dominates the thin film dynamics (i.e. $\sigma> A\Delta x$, \textcolor{black}{meaning that ${\cal N}_c< 1$}), no ripples are 
observed, whereas they appear whenever near-contact forces overcome the resistance to deformations due to surface tension \textcolor{black}{(${\cal N}_c\simeq 1$ or larger)}.
\textcolor{black}{Finally, if ${\cal N}_c\simeq 0$, droplets coalesce.}

\textcolor{black}{\subsection{Hydrodynamic effects: The role of the vorticity}}

The description of the physics of the film presented so far
\textcolor{black}{suggests that micro-recirculating fluid flows can considerably influence interface fluctuations,}
thus motivating a more detailed inspection of the hydrodynamic flow field within the film \cite{note}.

In Fig.\ref{flowfield} we show the velocity field for the case ${\cal N}_{c}>1$ (a) and ${\cal N}_{c}<1$ (b).
In the latter case, the fluid flow is almost everywhere unidirectional within both droplets, while it develops a couple of 
counter-rotating vortices, located at the periphery of the film, pushing the fluid outwards, thereby
facilitating the droplet approach. 
However, coalescence is inhibited because the flow in the bulk of the film is sufficiently intense to prevent 
interface contact. The flow displays a roughly uniform pattern, with intermittent bounces back and forth, due 
to the repulsive nature of the near-contact force, that mildly affects the interface.

When ${\cal N}_{c}>1$, the fluid flow within the film and in its surrounding is considerably more intense and
exhibits a more complex structure, highly chaotic in the middle and more regular at the sides, 
where, once again, two intense counter-rotating vortices are observed.
Such vortices significantly affect the dynamics of the fluid interface, which exhibits 
distinct bumps in the regions where the fluid flow is most intense.

\begin{figure}
  \includegraphics[width=1.0\linewidth]{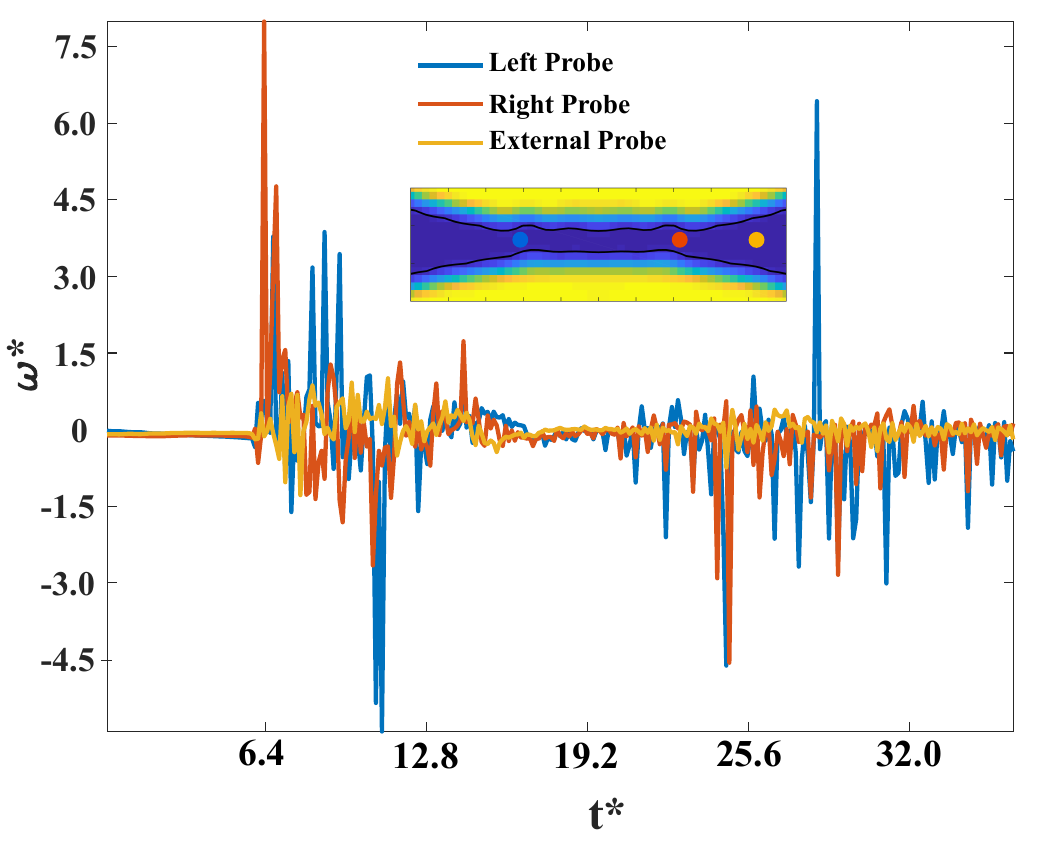}
\caption{Time evolution of the dimensionless vorticity $\omega^*=V_{\omega}/V_c$ \textcolor{black}{when ${\cal N}_c>1$}. 
The light yellow curve displays the time behaviour 
of $\omega^*$ for a probe located outside the film, while the blue and the orange ones 
refer to probes located at its extremities, on the left and on the right respectively. 
Positive and negative spikes denote clockwise and counterclockwise rotating vortices, respectively,
appearing at different stages  of the film evolution. \textcolor{black}{The typical fluid flow map is shown in Fig.\ref{flowfield}a,
at $t^*\simeq 6.6$ and $t^*\simeq 8$. The two instants were chosen in such a way to detect opposite sign vorticity at the location of the
two probes within the film.}}
\label{intermittent}
\end{figure}

\begin{figure}
\includegraphics[width=0.75\linewidth]{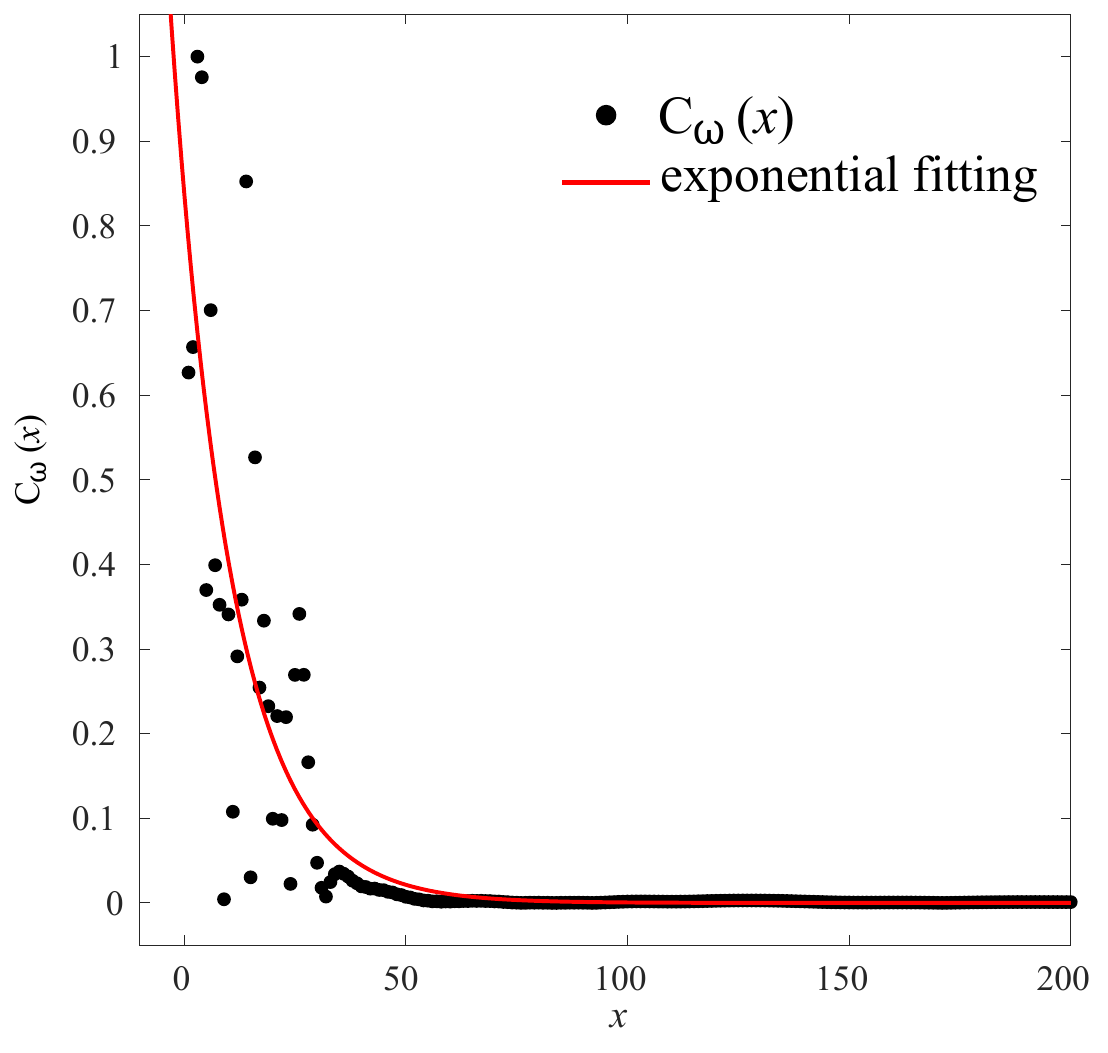}
\caption{Spatial correlation function of {\bf $\omega$}. Points represent the normalized spatial correlation function $C_{\omega}(x)$ of the vorticity ${\bf\omega}$. 
It is computed as $C_{\omega}(x)=|\langle\omega(x,t)\cdot\omega(x_c,t)\rangle|/|\langle\omega^2(x_c,t)\rangle|$, 
where $x_c$ is the centre of the film of fluid and $\langle...\rangle$ is a time average. 
$C_{\omega}(x)$ is fitted by a space decaying exponential function $\propto e^{-x/\xi_{\omega}}$, where $\xi_{\omega}\sim 15$.}
\label{fig4}
\end{figure}

Interestingly, the two lateral vortices are not stationary and their dynamic behaviour can be assessed 
by inspecting the vorticity $\boldsymbol{\omega}=\nabla\times{\bf v}$, ${\bf v}$ being the local velocity field. 
In Fig.\ref{intermittent}, we plot the time evolution of the dimensionless vorticity $\omega^*=V_{\omega}/V_{c}$
(where $V_{\omega}=\frac{\omega}{2} R$ and $\frac{\omega}{2}$ the angular velocity) in three distinct regions of the simulation box, two located at the two extremities of
the film (one left and one right) and the third outside. 
If the probe is located externally, $\omega^*$ fluctuates around zero, while, for the other two cases, $\omega^*$ 
displays intense spikes, both positive and negative, indicating the occurrence of clockwise (positive) 
or counterclockwise (negative) rotating vortices at different times.
The range of fluctuations of the interfaces can be evaluated by inspecting the vorticity space 
correlation function $C_{\omega}(x)=|\langle\omega(x,t)\cdot\omega(x_c,t)\rangle|/|\langle\omega^2(x_c,t)\rangle|$, where $x_c$ 
is the centre of the film of fluid and $\langle...\rangle$ is a time average. 
$C_{\omega}(x)$ can be fitted, to a reasonable accuracy, by a decaying exponential  
$\propto e^{-x/\xi_{\omega}}$, where $\xi_{\omega}\sim 15$ is the spatial correlation length (see Fig.\ref{fig4}). 
This value is approximately half of the length of the film ($\sim 35-40$ lattice sites), a range above which fluid 
vortices and interface fluctuations are expected to disappear.  


\textcolor{black}{\section{Conclusions}}
Summarising,  we have numerically investigated the effects induced by the NCIs on the fluid interface of two colliding droplets. 
It is found that the formation of interface ripples is controlled by the relative strength of NCI versus capillarity, measured by the
dimensionless number ${\cal N}_c$, which in turn triggers micro-vorticity patterns driving the interface undulations.
The latter exhibit a highly non-trivial dynamics, including persistent \textcolor{black}{fluctuations}, promoted by a complex 
flow field pattern,  which plays a crucial role on shape and stability of the intervening film of fluid between 
the droplets.  The thickness of the film at steady state \textcolor{black}{is in reasonable} agreement with experimental results. 
\textcolor{black}{Our results may be useful to suggest strategies for a more accurate design of}
soft flowing materials, whose size, structure and topology is likely to show significant dependence on the strength of near-contact interactions.

\section*{Acknowledgments}
A. T., A. M., M. L. and S. S. acknowledge funding from the European Research Council under the European Union's Horizon 2020 Framework
Programme (No. FP/2014-2020) ERC Grant Agreement No.739964 (COPMAT).

\end{document}